# Adaptive wavefront correction of dynamic multimode beam based on modal decomposition

**KUN XIE [4], WENGUANG LIU [*], QIONG ZHOU, LIANGJIN HUANG, ZONGFU JIANG, FENGJIE XI AND XIAOJUN XU**

[1]*College of Advanced Interdisciplinary Studies, National University of Defense Technology, Changsha, 410073, China*
[2]*State Key Laboratory of Pulsed Power Laser Technology, Changsha, 410073, China*
[3]*Hunan Provincial Key Laboratory of High Energy Laser Technology, Changsha, 410073, China*
[4]*xiekun279@gmail.com*
*lwg.kevin@163.com*

**Abstract:** We propose and demonstrate a method for the adaptive wavefront correction of dynamic multimode fiber beams for the first time. The wavefront of incident beam is reconstructed in real-time based on the complete modal information, which obtained by using the modal decomposition of correlation filter method. For the proof of principle, both of the modal decomposition and the wavefront correction are implemented using the same computer-generated hologram, which encoded into a phase-only spatial light modulator. We demonstrate the wavefront correction of dynamic multimode beam at a rate of 5Hz and achieve a 1.73-fold improvement on the average power-in-the-bucket. The experimental results indicate the feasibility of the real-time wavefront correction for the large mode area fiber laser by adaptive optics.

## 1. Introduction

Large mode area (LMA) fiber plays an essential role on the suppression of nonlinear effects in the high power fiber laser (HPFL) by enlarging the fundamental mode (FM) field area and lowering the power density in the core [1, 2]. As the core diameter increases, the number of supported propagating higher-order modes (HOM) in the fiber core will accordingly increase, which will cause a multimode (MM) operation and generally degrade the beam quality [3, 4]. Furthermore, the onset of transverse mode instabilities (TMI) will sharply degrade the beam quality when the output power exceeds a certain threshold, even if the initial beam quality is excellent and the HOM content is low [5, 6]. Thus, it appears that the key problem for LMA fiber is to enlarge the mode field area (MFA) and simultaneously realize effective single-mode operation. To solve this, many mode controlling approaches have been proposed based on the principle of increasing the HOM loss while reducing the FM loss. Among these approaches, bending is the most direct way, but it is not capable to discriminate the HOMs when the core diameter becomes larger than 20 μm [7] and TMI still occurs at high-power level [8, 9]. Novel fiber designs, such as large pitch fiber (LPF) [10], leakage channel fiber (LCF) [11], chirally-coupled-core (CCC) fiber [12], et al. have been presented to realize effective single-mode. However, these fibers should be critically designed, which greatly increases the difficulty of the manufacture technique. More importantly, none of these methods fully utilize the gain ability of the active fiber because HOMs are mostly lost, which is indeed confined the output power to merely single mode (SM). From another perspective, if the MM beam output could be allowed on the premise that its beam quality could be reformed, it will significantly unleash the gain potential and simultaneously the beam quality is also preserved.

Actually, the wavefront correction of a static single HOM has been demonstrated by using the binary phase plate [13-16] or the interferometric element [17]. However, these rigid devices are not capable to correct the dynamic wavefront of the beam with more than one HOM. Comparing with that, the adaptive optics (AO) technique is more competent, which has successfully achieved the dynamic correction of the wavefront of the high power solid-state lasers [18-20]. Generally, in the AO systems, there are two ways for the close-loop control. One is based on the iterative optimization algorithm [21-25] and the other one is based on the wavefront conjugation [26]. However, both of them have some deficiencies in the application of MM fiber laser. For the first one, the iteration process of optimization algorithm often takes more than hundreds of steps and several minutes due to the large number of optimization parameters, thus the previous experimental demonstrations have only involved static MM beams [21-24]. For the second one, current wavefront sensors (WFS) (such as the Shack-Hartmann wavefront sensor (SHS) and lateral shearing interferometer (LSI) [27]) are insufficient to accurately measure the phase steps in the wavefront of MM beams [25, 28-30]. To the best of our knowledge, the dynamic correction of the MM beam has not been reported.

In this paper, we proposed a method for the wavefront correction of dynamic MM beam. In the method, the wavefront of incident beam is reconstructed based on the modal decomposition (MD) technique. The MD and wavefront correction are implemented with a set of transmission functions, which is encoded into a computer generated hologram (CGH). The effectiveness of this method is also experimentally verified by static and dynamic MM beams respectively. The rest of this paper is formatted as follows. Section 2 details the methods of the MD and wavefront correction. The experimental setup and results are presented in Section 3. Conclusions are given in Section 4.

## 2. Method

### 2.1 Modal decomposition and wavefront reconstruction

In our system, the wavefront measurement of MM beam is achieved by means of MD technique. With the prior information of eigenmode fields and the modal coefficients (weight and phase) determined by MD, the wavefront of incident beam could be precisely reconstructed. This wavefront measurement method can not only accurately measure the phase steps, but also achieves the same rate as MD, which paves the way to measure the wavefront of MM fiber beam in real-time. In the past few years, several MD methods have been proposed, such as the numerical analysis method [31-33], spatially and spectrally resolved imaging ($S^2$) technique [34, 35], ring-resonators method [36], and wavefront analysis method [28, 37]. In this paper, correlation filter method (CFM) [38-42] is adopted to MD because it only requires straightforward algebraic calculation to obtain the modal components, which significantly increase the rate of in-line MD up to 30Hz [43] (the maximum rate with other in-line MD methods is 9Hz [31]) and greatly benefits real-time operation.

The principle of MD with CFM will be introduced in this section. The electric field propagating in the fiber can be described by the superposition of a complete set of eigenmodes. In weakly guiding fibers, these modes can be well approximately represented by the linearly polarized (LP) mode

$$U_r(\mathbf{r}) = \sum_{n=1}^{N} c_n \cdot \psi_n(\mathbf{r}), \qquad c_n = \rho_n \cdot e^{i\phi_n} \tag{1}$$

where $U_r(\mathbf{r})$ is the complex amplitude of optical field in each polarization direction, $\psi_n(\mathbf{r})$ is the normalized electric field distribution of the $n^{th}$ LP mode guiding in the fiber, $\rho_n$ is the modal amplitude and $\phi_n$ is the intermodal phase difference, $c_n = \rho_n \cdot e^{i\phi_n}$ is the coefficient of each mode. The total number of the modes is $N$. For a fiber with given parameters, the $\psi_n(\mathbf{r})$ of guiding modes can be calculated in advance. Thus, any field propagating in this fiber could be characterized by a one-dimensional set of $c_n$. The purpose of MD is to analyze the actual set of $c_n$.

In the CFM, a correlation filter encoded in the CGH is the most crucial element. The beam emitted from the end face of fiber is diffracted by the CGH and then Fourier transformed by a lens. The far-field on the optical axis is proportional to the correlation of the incident optical field and the transmission function of the CGH. In this way, the modal amplitudes and phases can be measured by designing the CGH with a set of specific transmission functions [38, 44]. For the modal amplitude measurement, the transmission function of each mode is

$$T_n(\mathbf{r}) = \psi_n^*(\mathbf{r}), n = 1, 2, ..., N \tag{2}$$

Using this transmission function, the intensity on the optical axis in the Fourier plane indicates the modal weight, which can be calculated by

$$\rho_n^2 \propto \left| \mathbb{F}\{U(\mathbf{r})T_n(\mathbf{r})\} \right|^2 = I(0,0) \tag{3}$$

where $I(x, y)$ is the intensity distribution in the Fourier plane. For the modal phase measurement, the FM is chosen as the reference mode. Then, two transmission functions of each mode are given as follows

$$T_n^{\cos}(\mathbf{r}) = \left[\psi_0^*(\mathbf{r}) + \psi_n^*(\mathbf{r})\right]/\sqrt{2}, \quad T_n^{\sin}(\mathbf{r}) = \left[\psi_0^*(\mathbf{r}) + i\psi_n^*(\mathbf{r})\right]/\sqrt{2}, \, n = 2, 3, ..., N \tag{4}$$

Here the diffracted field of each transmission function carries information of the FM and the $n^{th}$ mode, thus its far-field diffraction pattern represents the intermodal interference of them. The intermodal phase can be calculated by

$$\phi_n = \begin{cases} -\arctan\left[\dfrac{2I_n^{\sin} - \rho_n^2 - \rho_0^2}{2I_n^{\cos} - \rho_n^2 - \rho_0^2}\right] &, \left(2I_n^{\cos} - \rho_n^2 - \rho_0^2\right) \geq 0 \\ \pi - \arctan\left[\dfrac{2I_n^{\sin} - \rho_n^2 - \rho_0^2}{2I_n^{\cos} - \rho_n^2 - \rho_0^2}\right] &, \left(2I_n^{\cos} - \rho_n^2 - \rho_0^2\right) < 0 \end{cases} \tag{5}$$

where $I_n^{\sin}$ and $I_n^{\cos}$ represent the far-field intensity on the optical axis of the diffracted field of $T_n^{\sin}(\mathbf{r})$ and $T_n^{\cos}(\mathbf{r})$, respectively. Using these transmission functions above, all modal amplitudes and phases can be obtained.

For the MD involving $N$ modes, $N + 2(N-1) = 3N - 2$ transmission functions are required. These transmission functions are encoded in a CGH by means of angular multiplexing. Each transmission function $T_n(\mathbf{r})$ is modulated onto a spatial carrier frequency $K_n(\mathbf{r})$ and then superimposed together to obtain the final transmission function as follows

$$T_{Measure}(\mathbf{r}) = \sum_{n=1}^{N} T_n(\mathbf{r})e^{iK_n\mathbf{r}} + \sum_{n=2}^{N} \left(T_n^{\cos}(\mathbf{r})e^{iK_n^{\cos}\mathbf{r}} + T_n^{\sin}(\mathbf{r})e^{iK_n^{\sin}\mathbf{r}}\right) \tag{6}$$

In the Fourier plane, the diffraction patterns corresponding to their respective transmission functions are simultaneously arranged at different positions.

Based on the measured modal amplitudes $\rho_n$ and phases $\phi_n$, the complex amplitude of the optical field in the investigated polarization direction could be reconstructed according to Eq. (1). To evaluate the agreement of the actual field and the reconstructed field, the cross correlation function is proposed

$$C = \left| \frac{\iint \Delta I_{Rec}(x,y) \Delta I_{Mea}(x,y) dxdy}{\sqrt{\iint \Delta I_{Rec}^2(x,y) dxdy \iint \Delta I_{Mea}^2(x,y) dxdy}} \right| \tag{7}$$

where $\Delta I_j(x,y) = I_j(x,y) - \overline{I_j(x,y)}$, $\overline{I_j(x,y)}$ is the mean value of the near-field intensity profile with $j = Rec, Mea$ denoting the reconstructed and measured one respectively. For an ideal MD and reconstructed field, the correlation coefficient is one; for an inaccurate result, it's close to zero.

## 2.2 Adaptive wavefront correction

From the former section, it is shown that the phase distribution $\phi_{Rec}(\mathbf{r})$ of the optical field can be obtained from the complex amplitude reconstructed by MD. Thus, the MD can be used as the WFS in the AO system. For a proof of concept demonstration, our aim is to verify the feasibility of real-time wavefront correction, rather than actually correct the entire optical field. Thus, it is no need to use an individual wavefront corrector, such as deformable mirrors (DM) or phase-only spatial light modulators (SLM). In our system, the corrector is substituted by a transmission function, which carries the conjugate phase of the incident optical field, as shown below

$$T_{Correct}(\mathbf{r}) = e^{-i\phi_{Rec}(\mathbf{r})} \cdot e^{iK_{Correct}\mathbf{r}} \tag{8}$$

The diffraction pattern of this transmission function can also be observed on the Fourier plane, which is equivalent to the far-field intensity profile of the corrected beam. Besides, another mirror-like transmission function with a tilt phase only is added to monitor the original far-field intensity profile without correction, as shown below

$$T_{Ori}(\mathbf{r}) = e^{iK_{Ori}\mathbf{r}} \tag{9}$$

The transmission functions in Eq. (8) and Eq. (9) act as a wavefront corrector and a beam splitter respectively. This method not only simplifies the experimental configuration, but also allows the acquirement of the two far-field intensity profiles with and without correction in the same optical path and camera simultaneously.

In the real-time correction procedure, modal information is measured with the $k^{th}$ CGH firstly, then the $(k+1)^{th}$ CGH is generated with conjugate phase of the reconstructed optical field. The final transmission function of the CGH is a superposition of the stationary measurement transmission function and the dynamic correction transmission function, which can be expressed as

$$T_{Final}(\mathbf{r}, k+1) = T_{Measure}(\mathbf{r}) + A_{Ori} \cdot T_{Ori}(\mathbf{r}) + A_{Correct} \cdot T_{Correct}(\mathbf{r}, k) \tag{10}$$

The $A_{Correct}$ and $A_{Ori}$ can be adjusted to obtain the proper brightness of the far-field intensity profiles with and without correction, respectively. In the Fourier plane, the mode information and the far-field intensity profiles are acquired simultaneously by one camera. By selecting a series of carrier frequencies, the diffraction patterns can be spatially separated without crosstalk. To illustrate the arrangement of diffraction patterns, a simulation result generated with a typical MM beam is shown in Fig. 1. The +1$^{st}$ Rank diffraction pattern on the Fourier plane is shown in Fig. 1(a), the spots for modal weight analysis and modal phase analysis are in the red polygon and yellow polygon respectively; the far-field intensity profiles are in the white box. The modal weight and the near-field intensity profile of the beam used in the simulation are shown in Fig. 1(b), (c) respectively.

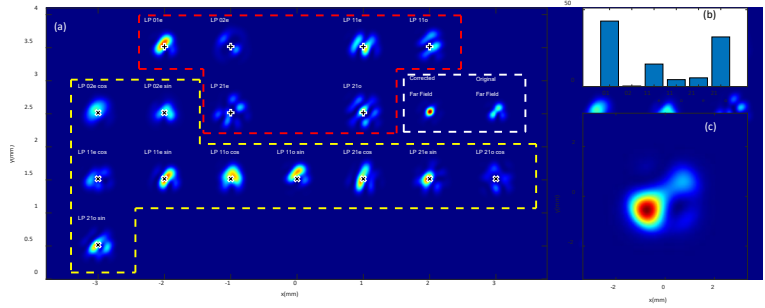

Fig. 1. Simulation of the diffraction pattern of a MM beam. (a) +1st Rank diffraction pattern on the Fourier plane. (b) Modal weight of the incident beam. (c) Near-field intensity profile of the incident beam. The spots in the red polygon are for modal weight analysis; the spots in the yellow polygon are for modal phase analysis; the spots in the white box are for far-field monitoring.

Fig. 2 shows the far-field intensity profiles without and with wavefront correction of the six eigenmodes of a 0.065-NA LMA fiber with core of 25μm (the operation wavelength is 1064nm; the near-field is expended 125 times and the far-field is obtained by a focus lens of f=175mm). As expected, with the wavefront correction, each of the far-field intensity profile of MM beam transforms from a spot with multiple lobes to a Gaussian-like spot. The power-in-the-bucket (PIB) of far-field intensity profile and beam propagation factor $M^2$-parameter are used to evaluate the beam quality respectively [45]. The diameter of the bucket is 97μm, which is defined at the $1/e^2$ of the peak intensity of the FM. With the correction, the PIBs of $LP_{11}$, $LP_{21}$ and $LP_{02}$ mode are increased from 50.9%, 23.6% and 26.3% to 74.5%, 68.6% and 83.1%, respectively. In contrast, the $M^2$-parameter of each mode remains the same with the correction. This is because the far-field beam width defined by the second-order moment exaggerates the contribution of the outer wings of the intensity profile [45, 46]. It also can be seen from the PIB curves shown in Fig. 2, the corrected curve is building up faster than the original one in the range of radius < 40μm, and contains approximately 70% of the power. Since the AO system is mainly used to improve the far-field power density of the output laser, the PIB should be more appropriate to evaluate the beam quality in this paper.

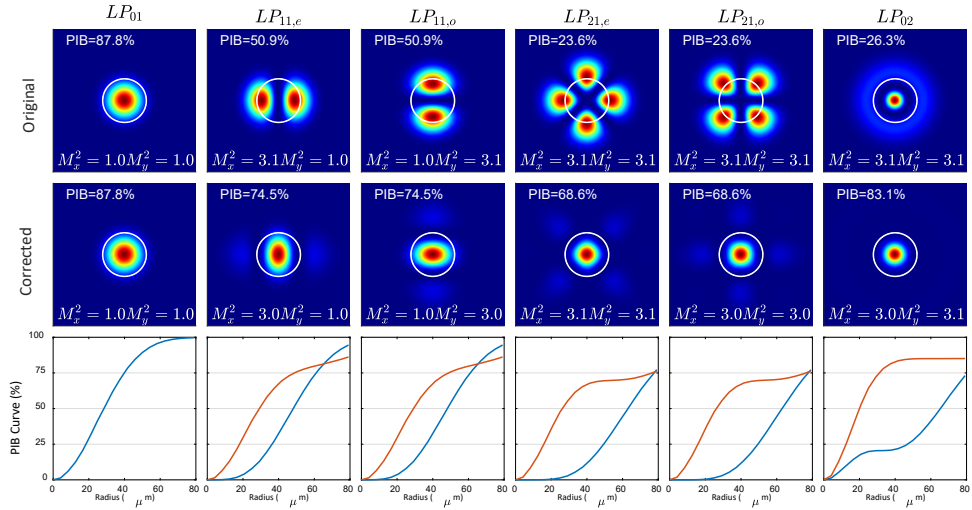

Fig. 2. Simulated comparison of the far-field intensity profiles of six LP modes. The subgraphs in the first row indicate the original far-field intensity profiles without correction. The subgraphs in the second row indicate the corrected far-field intensity profiles. The white circle in each subgraph indicates a D=97μm bucket. Subgraphs in the third row indicate the PIB vs the bucket radius. The blue curve and red curve indicate the original and corrected PIB respectively.

## 3. Experiments and results

### 3.1 Experimental setup

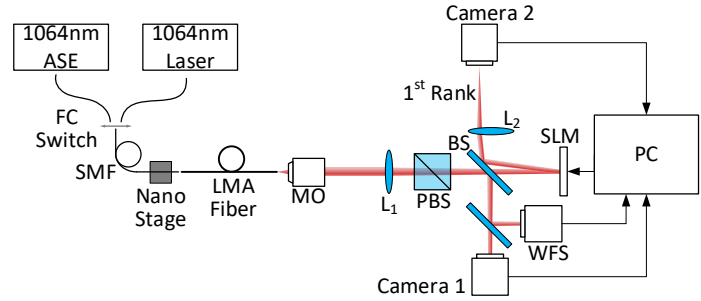

Fig. 3 Experimental setup. SMF, single-mode fiber; FC, fiber connector; LMA fiber, large-mode-area fiber; L$_1$, L$_2$, lenses; MO, microscopic objective; BS, beam splitter; PBS, polarizing beam splitter; SLM, spatial light modulator; WFS, Shack-Hartman wavefront sensor; 1$^{st}$ Rank, +1$^{st}$ rank diffraction beam of CGH.

The experimental setup as schematized in Fig. 3 consists of a beam source part to generate the MM beam with aberrated wavefront, and a wavefront correction part. The beam source part comprises two light sources, a single-mode fiber (SMF) pigtailed single-frequency laser at 1064 nm serves as a coherent source to excite mixed modes in the LMA fiber, and an ASE source at 1064 nm with 10 nm bandwidth serves as an incoherent source to align and calibrate the MD system. These two light sources, respectively, connect to a SMF with a fiber connector, which allows rapid switch between them. Then the beam from the SMF is free-space coupled into a step-index LMA fiber with core diameter of 25 μm and core NA of 0.065. The supported eigenmodes in this LMA fiber at operating wavelength of 1064 nm are $LP_{01}$, $LP_{11e, o}$, $LP_{21e, o}$, and $LP_{02}$ mode. By moving the output end face of the SMF with a 3-axis nano-position stage, the coupling condition between the SMF and the LMA fiber can be changed, thus exciting variable MM beams in the LMA fiber. For the output beam of the LMA fiber, a 4-f system with magnification factor of 125 is used to get an enlarged optical field of the fiber end face. The 4-f system consists of a microscopic objective and an achromatic doublet lens whose focal lengths are 4 mm and 500 mm respectively (500/4=125). This enlarged field images at a phase-only reflective SLM (1920×1080 pixels of 6.4 μm pitch) and a near-field camera (Camera 1, 1920×1200 pixels of 5.86 μm pitch) through a beam splitter (BS). For comparison, an SHS (21×21 subapertures of 150 μm pitch) is placed in the position symmetrical with the Camera 1 to measure the wavefront. The CGH introduced in Section 3 is displayed on the SLM for the MD and the wavefront correction. The field is diffracted by the CGH and then Fourier transformed by a lens (Lens 2, f=175mm). A camera (Camera 2, 1928×1448 pixels of 3.69 μm pitch) is placed at the Fourier plane of the SLM to acquire the +1$^{st}$ rank diffraction pattern which contains the correlation information and the far-field intensity profiles with and without correction. Since the SLM has a specific working polarization direction, a polarizing beam splitter (PBS) is used to choose one state of polarization of the optical field. The MD and the wavefront correction were only proceeding in the state of polarization parallel to the macro-axis of the SLM.

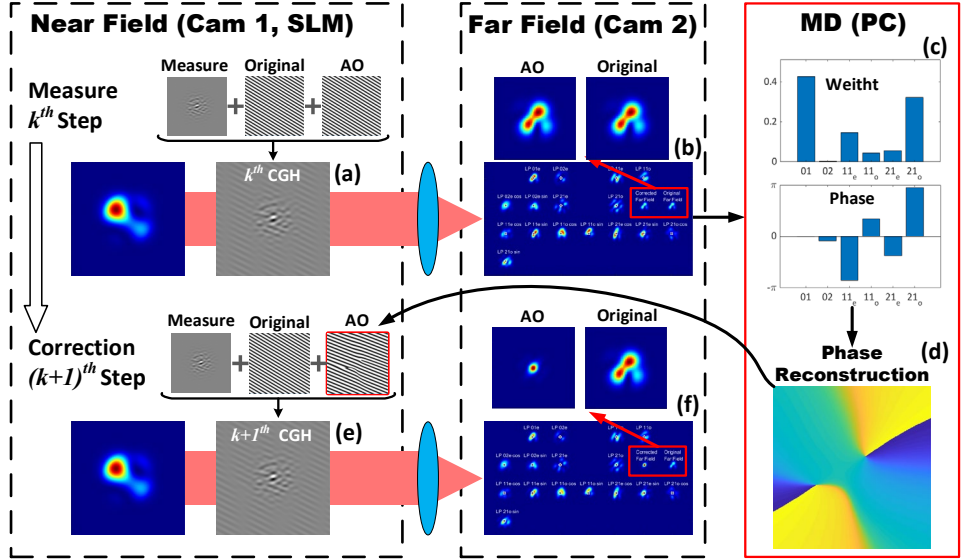

Fig. 4. Workflow of the adaptive wavefront correction in the experiment. (a) The $k^{th}$ CGH displayed on the SLM; (b) the diffraction pattern in the $k^{th}$ frame of the Camera 2; (c) the modal components analyzed by MD; (d) the reconstructed near-field phase distribution; (e) the $(k+1)^{th}$ CGH with conjugated phase of reconstructed field; (f) the far-field intensity profiles acquired from the diffraction pattern in the $(k+1)^{th}$ frame of the Camera 2.

To give a detailed illustration of the experiment workflow, a typical simulation result is given in Fig. 4 as a paradigm. The $k^{th}$ CGH (Fig. 4(a)) is firstly generated with the $T_{Correct}$ of a tilt phase only and displayed on the SLM. By analyzing the diffraction patterns in the $k^{th}$ frame (Fig. 4(b)) with MD algorithm, the modal components (Fig. 4(c)) in the fiber could be obtained and then the original near field (Fig. 4(d)) could be reconstructed in real-time. Next, according to Eq. (10), the $(k+1)^{th}$ CGH which contains the conjugated phase of the near field (Fig. 4(e)), will be generated and sent to SLM. After that, the corrected far-field intensity profile can be acquired from the $(k+1)^{th}$ frame of Camera 2 (Fig. 4(f)).

### 3.2 Results and discussion

The experimental setup described above was used to demonstrate the wavefront correction for the MM beam. Its wavefront correction capability was verified under static and dynamic conditions separately.

In the first experiment, we excited a series of static MM beams by randomly coupling the SMF and LMA fiber with slight misalignment. Here the details of a typical case is illustrated in Fig. 5. The measured near field intensity profile is composed of one main lobe and several side lobes (Fig. 5(b)). According to the modal weight spectrum obtained by MD (Fig. 5(a)), the beam mainly consists of 40.6% $LP_{21}$ odd mode, 29.0% $LP_{11,o}$ mode and 12.2% $LP_{01}$ mode.

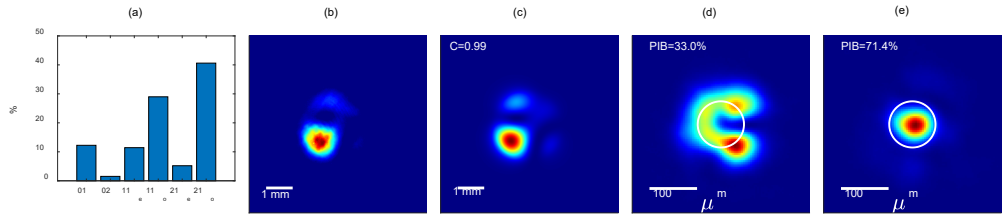

Fig. 5. Typical example of the static correction. (a) The modal weight spectrum. (b) The measured near-field intensity profile. (c) The reconstructed near-field intensity profile. (d) The original far-field intensity profile. (e) The corrected far-field intensity profile. The white circle in each subgraph indicates a D=97μm bucket.

The reconstructed near-field intensity profile in Fig. 5(c) shows a high consistency with the measured one. The cross correlation C is 0.99, which suggests a high reliability of the MD. The far-field intensity profiles without and with correction are shown in Fig. 5(d) and (e). Due to the phase fluctuation caused by the HOMs, the original far-field intensity profile is divided into several lobes. With the correction, the far-field intensity profile transforms to a Gaussian like distribution. The PIB of the corrected beam increases from 33.0% to 71.4%, which is close to 87.8% of the FM.

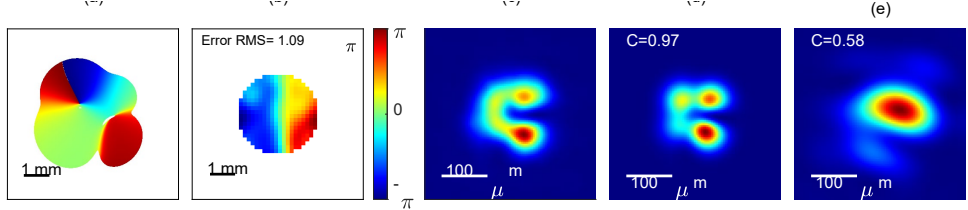

Fig. 6. Comparison of the MD and the SHS reconstructed wavefront. (a) The wavefront reconstructed by MD. (b) The wavefront reconstructed by SHS. (c) The measured far-field intensity profile. (d) The calculated far-field intensity profile with wavefront reconstructed by MD. (e) The calculated far-field intensity profile with wavefront reconstructed by SHS.

In order to compare the wavefront measurement capability of MD and SHS, the wavefronts reconstructed by MD and SHS are shown in Fig. 6(a), (b) respectively. It shows that the wavefronts reconstructed by two means are quite different and the root-mean-square (RMS) difference between them reaches 1.09 π. Then, with the two reconstructed wavefronts, we respectively constructed two near-field optical fields using the measured near-field intensity profile, and calculated their far-field intensity profiles according to the Fraunhofer diffraction. The far-field intensity profile calculated based on the MD wavefront (Fig. 6(d)) is almost consistent with the measured profile (Fig. 6(c)), and the correlation coefficient C is up to 0.97. This result reveals the high accuracy of the wavefront reconstruction with MD. In contrast, the far-field intensity profile calculated based on the SHS wavefront (Fig. 6 (e)) shows obvious difference to the measured one, and the correlation coefficient C is only 0.58. This wide deviance is actually caused by the SHS's insufficient capability to measure the 0-π phase steps in the wavefront, which appear in the low-intensity areas between the lobes, as mentioned in Section 1. Therefore, for the MM beam, reconstructing its wavefront by MD is more accurate than using SHS.

Based on the experimental verification with static field, the adaptive wavefront correction based on MD was testified with dynamic field then. By randomly adjust the nano-position stage, the modal components in LMA fiber can be varied. We excited MM beam with ever changing modal components to verify the effectiveness of the correction under dynamic condition. The far-field PIB of the original beam and the corrected beam are shown in Fig. 7 by the blue and red curves respectively, and the modal weight of FM is represented by the black dashed curve. It can be found from Fig. 7 that for the beam without correction, the far-field PIB is mainly relative to the modal weight of FM. A lower FM weight leads to a sharp drop in the PIB. For the beam with correction, the mean value of PIB increases from 40.1% to 69.3% and the minimum value of PIB is always above 54%, which suggests a significant improvement in the beam quality. In other words, in order to achieve the same performance, the output power of the MM HPFL only needs to be 1.11 times that of the SM HPFL, considering that the PIB of the $LP_{01}$ mode is 87.8% (87.8%/69.3%=1.11). To highlight the real-time correction ability of our method, the correction process of Fig. 7 is presented in the Visualization 1.

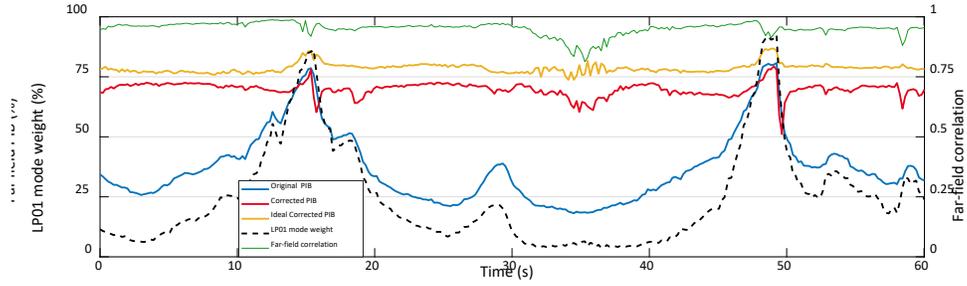

Fig. 7. The far field intensity PIB vs time curves for the original beam without correction (blue) referred to the corrected beam (red). The black dashed curve represents the modal weight of FM. The yellow curve represents the PIB of the ideal corrected beam. The green curve represents the cross correlation between the reconstructed far-field intensity profile and the measured one. The near-field intensity profiles of correspond time are shown on the top of figure.

The simulated result of ideal correction is obtained by substituting the phase of reconstructed field with a perfect plane phase, and the PIB of its far-field profile is represented by the yellow curve in Fig. 7, with an average value of 79.2%. It also can be seen from Fig. 7 that the experimental result of PIB (see the red curve in Fig. 7) is lower than the simulated ideal result, which is mainly caused by the aberrations and noise. Firstly, a few aberrations are introduced by the imperfection of Fourier lens $L_2$. Then the quality of far-field diffraction pattern will be degraded by the aberrations, which introduces errors into the MD and the calculation of PIB. To eliminate the influence of these aberrations, the CGH should be superposed on a correction phase which could compensate these aberrations. Secondly, the signal-to-noise ratio (SNR) of diffraction pattern images are affected by noise. The useless $0^{th}$ rank of diffraction pattern is much brighter than the $+1^{st}$ rank which we measured. Thus both of the outer-wings and stray light of the $0^{th}$ order diffraction pattern introduce background noise to the $+1^{st}$ order diffraction pattern, which also causes the decrease of the PIB. Besides, it can be seen that the cross correlation between the reconstructed far-field intensity profile and the measured one (see the green curve in Fig. 7) decreases with the reduction of the FM modal weight (see the black dashed curve in Fig. 7). This is because the LMA fiber was bent in the experiment, which introduced a little bit higher loss into the HOMs. Thus, the power of the output beam decreases as the FM modal weight decreases, and the far field diffraction pattern becomes faint (see the subfigure of T=34.1s in the Fig. 7). Since the exposure configuration of the Camera 2 is fixed in the experiment, the SNR of diffraction pattern is decreased, which leads to more errors of the reconstruction of optical field and thus reduces the effect of wavefront correction. Therefore, for some cases where the FM modal weight is extremely low and the HOM dominate the beam (30~40s in Fig. 7), the far-field PIB of the corrected beam shows a slightly decrease. However, this shortcoming could be overcome by adaptively controlling the exposure time and gain of camera to preserve the SNR of diffraction pattern.

In our current system, the rate of MD and wavefront reconstruction is 25Hz, which is limited by the maximum frame rate of Camera 2. The rate of close-loop correction is only 5 Hz, due to the calculation time of the high-resolution CGH and the refresh time of SLM. However, the rate of MD could be further improved by replacing the camera with a fast photodiode array and implementing the MD algorithm via a field-programmable gate array (FPGA). In addition, substituting the SLM by a DM with dozens of actuators as the corrector could achieve a correction rate up to ~5kHz [47]. This method suggests great potential for correcting the drastically fluctuating wavefront of the MM beam. This makes it possible to obtain a high far-field PIB in the high power LMA fiber laser even when TMI occurs.

## 4. Conclusion

In conclusion, we presented the adaptive wavefront correction in a LMA fiber based on the MD for the first time. Experimentally, we shown that the correlation filter method could decompose the optical field in a six-modes LMA fiber with high precision. The close-loop wavefront correction for the dynamic MM beam was demonstrated at a rate of 5Hz. With the wavefront correction, the mean value of the far-field PIB of the dynamic MM beam increased from 40% to about 70%, which is close to the performance of a quasi-SM beam. Accordingly, we believe this adaptive wavefront correction method will be an effective tool to combat the beam quality degradation caused by TMI in the high power LMA fiber laser by increasing the far-field PIB close to a SM level. In order to increase the brightness of high power LMA fiber laser actually, we expect to construct an AO system with a deformable mirror in the future and correct the wavefront of the entire field.

## Funding

Natural Science Foundation of China (NSFC, No.11504423)